# DIAGNOSTICS FOR LINAC OPTIMIZATION WITH MACHINE LEARNING

R. Sharankova*, M. Mwaniki, K. Seiya, M. Wesley
Fermi National Accelerator Laboratory, Batavia, IL 60510, USA

*Abstract*

The Fermilab Linac delivers 400 MeV H- beam to the rest of the accelerator chain. Providing stable intensity, energy, and emittance is key since it directly affects downstream machines. To counter fluctuations of Linac output due to various effects to be described below we are working on implementing dynamic longitudinal parameter optimization based on Machine Learning (ML). As inputs for the ML model, signals from beam diagnostics have to be well understood and reliable. In this paper we discuss the status and plans for ML-based optimization as well as preliminary results of diagnostics studies.

## THE FERMILAB LINAC

The Fermi National Accelerator Laboratory (Fermilab) Linac accelerates H- beam from 750 keV to 400 MeV. The Linac is preceded by the Pre-Accelerator comprising the Ion Source, the Low Energy Beam Transport (LEBT) line, a radio-frequency quadrupole (RFQ), and the Medium Energy Transport (MEBT) line. The magnetron ion source ionizes hydrogen gas into plasma, then extracts and accelerates a beam of negative H ions. The continuous H- beam is chopped, bunched and accelerated through the rest of the Pre-Accelerator from 35 keV to 750 keV of kinetic energy before entering the Linac. The Linac comprises three parts: a Drift Tube Linac (DTL), a transition section and a Side Coupled Linac (SCL). The DTL is composed of 207 drift tubes spread across 5 tanks. It operates at RF frequency of 201.25 MHz and accelerates beam to 116.5 MeV. The SCL has 7 modules, operating at resonant frequency of 805 MHz and accelerates beam to 401.5 MeV. A buncher and a vernier cavity located in the transition section allow for longitudinal matching between the DTL and the SCL. During regular operations, the Fermilab Linac has an output of roughly 25 mA and pulse length of 35 $\mu$s, with transition efficiency $\geq$ 92%

## LINAC RF & LLRF

DTL RF field amplitude is controlled by the Marx modulator logic controller which in turn controls the 5 MW power tube modulator voltage [1]. The RF phase is controlled by the low level RF (LLRF) module in a VME eXtension for Instrumentation (VXI) crate. Each SCL module is powered by a 12 MW klystron with VXI based LLRF phase and amplitude control.The amplitude and phase are set via the Fermilab control network (ACNET) [2] which sends desired settings to the front-end card in the LLRF VXI crate. In total there are 34 RF parameters (17 phase set points and 17 field gradients) that can be manipulated to affect the overall longitudinal accelerating field in the Linac.

## LINAC DAILY TUNING

Stable Linac output is crucial for downstream machines. Ambient temperature and humidity variations are known to affect resonance frequency of the accelerating cavities which induces emittance growth and increased particle loss. In addition, the energy and phase space distribution of particles emerging from the ion source are subject to fluctuations. To counter such effects, operators perform daily tuning. This tuning consists of hand-scanning a handful of RF parameters and trying to maximize beam currents while minimizing losses along the Linac. Figure 1 shows an example of a scan of the RFQ RF phase set point and its effect on total Linac losses and beam currents.

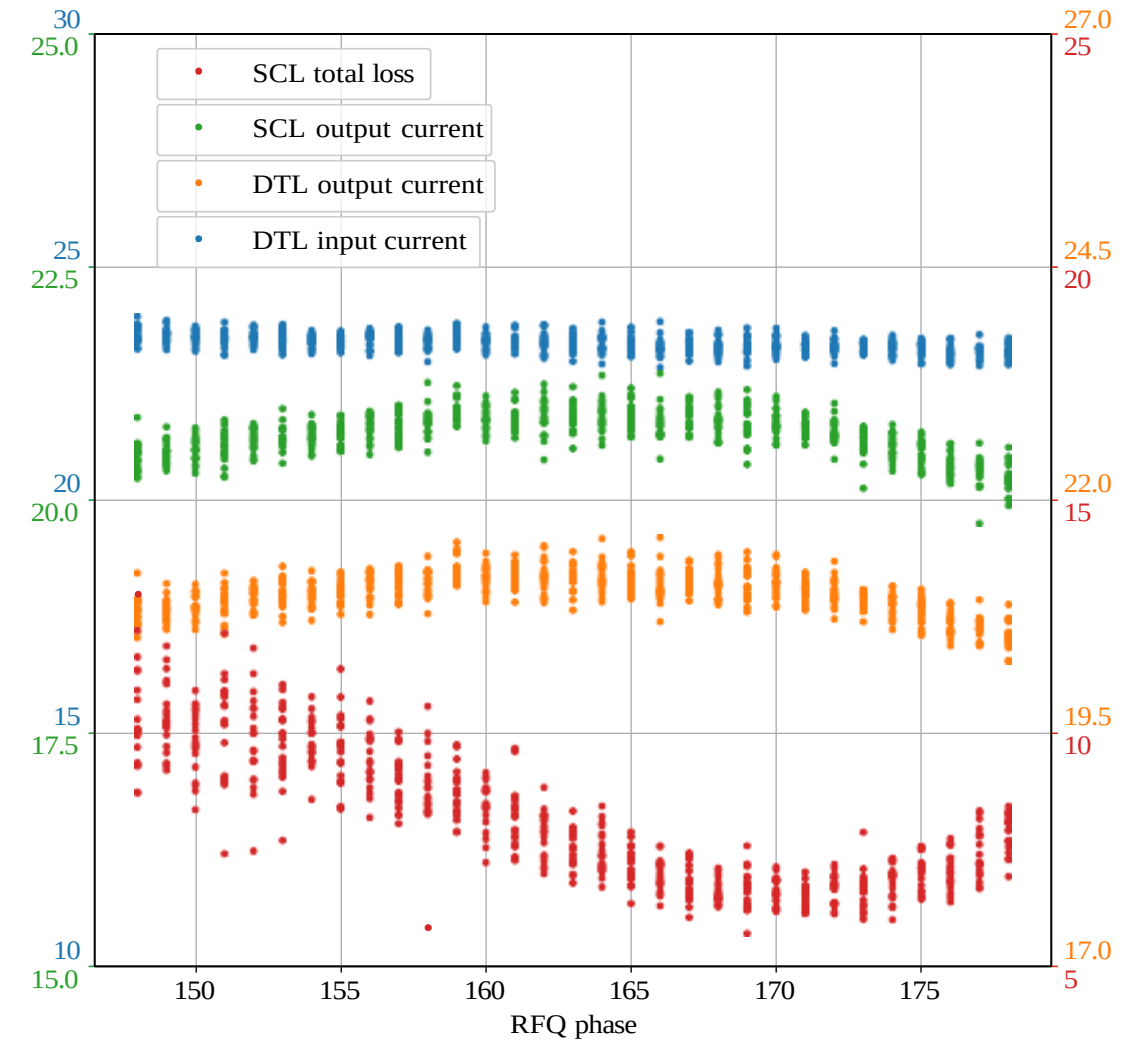

Figure 1: Effect of RFQ phase set point on total Linac beam loss (red) and output beam current (green). Also showing Linac input (blue) and DTL output currents (orange).

## RF OPTIMIZATION WITH ML

The hand-tuning procedure outlined above faces several challenges. Human operators cannot optimize in multi-parameter space simultaneously, so there is always the possibility of being off-optimal for the first N-1 devices after tuning device N. Additionally hand-tuning is done when personnel is available, and not necessarily whenever the Linac conditions change.

* rshara01@fnal.gov

To resolve these challenges, we are working towards automating the tuning procedure, and we are exploring ML applications for that goal. Our approach has been to train a Deep Neural Network (DNN) to recognize the underlying correlations between observed diagnostic data and Linac RF parameters. The plan is to use the network predictions in a control scheme which adjusts phase set points to change the Linac state (as defined by diagnostic readings) back to a desired condition. The choice of DNN among all available ML algorithms was driven by the fact that a network with enough nodes is able to approximate almost any analytical model, without having to know the exact underlying correlations between observables.

Originally we mirrored the hand-tuning procedure and tried to train a network using loss monitor and toroid data. This was unsuccessful, and a deeper look at the data revealed losses and beam currents are very susceptible to external factors and largely fluctuate even under the same RF setting. We concluded they are not the best data for training.

In out second attempt used Beam Position Monitor (BPM) data for network training. BPMs measure beam horizontal and vertical positions, beam phase w.r.t. each cavity's RF reference signal, and beam intensity. They are installed at the beginning and the end of most RF cavities, 6 in total in the DTL and 28 in the SCL. These give us information about both the longitudinal and transverse state of the beam.

The network discussed here is a fully-connected feed-forward NN consisting of an input layer, a normalization layer which scales all inputs to the same range, four hidden layers with 100 nodes each, and an output layer. Network was implemented in Tensorflow framework [3]. Network hyperparameters and architecture were loosely optimized. Using all BPMs from the DTL as well as BPMs from Module 6 in SCL as inputs, we trained the DNN to predict the change in RFQ, Buncher and Tank 5 phase set points w.r.t. a fixed reference. The resulting prediction resolution as estimated on a test sample from the same time period as the training data (May 2022) is shown in Fig. 2 (top). The network is able to predict the RF phases with precision (standard deviation of Gaussian fit) of 2.50, 1.54 and 0.11 degrees respectively for the RFQ, Buncher and Tank 5. This is on-par with precision achievable in hand-tuning.When we attempted to apply the model to data from 3 weeks later, however, both prediction bias and resolution were much worse, see Fig. 2 (bottom).

Further investigation showed that BPM data drifts throughout the day, sometimes starting even upstream of the Linac. Figure 3 shows the evolution of BPM phase (left) and horizontal and vertical positions (right) during the period from 21 May 2022 1AM to 10AM. The first hour of the day is used as reference. The reference measurements are then subtracted from each subsequent hourly data set, and the residuals are plotted. The data is ordered horizontally corresponding to BPM placement along the Linac, starting with the BPM at end of DTL Tank 2 (labels L:BPM2OF/L:BPV2OT), and ending with the BPM at the end of SCL Module 7 (L:D74BF/L:D74BPV). During this period there was no change to Linac RF parameters, however we observe changes in beam transverse position as upstream as the first Linac BPM. One possible interpretation is a change in the energy coming out of the ion source.

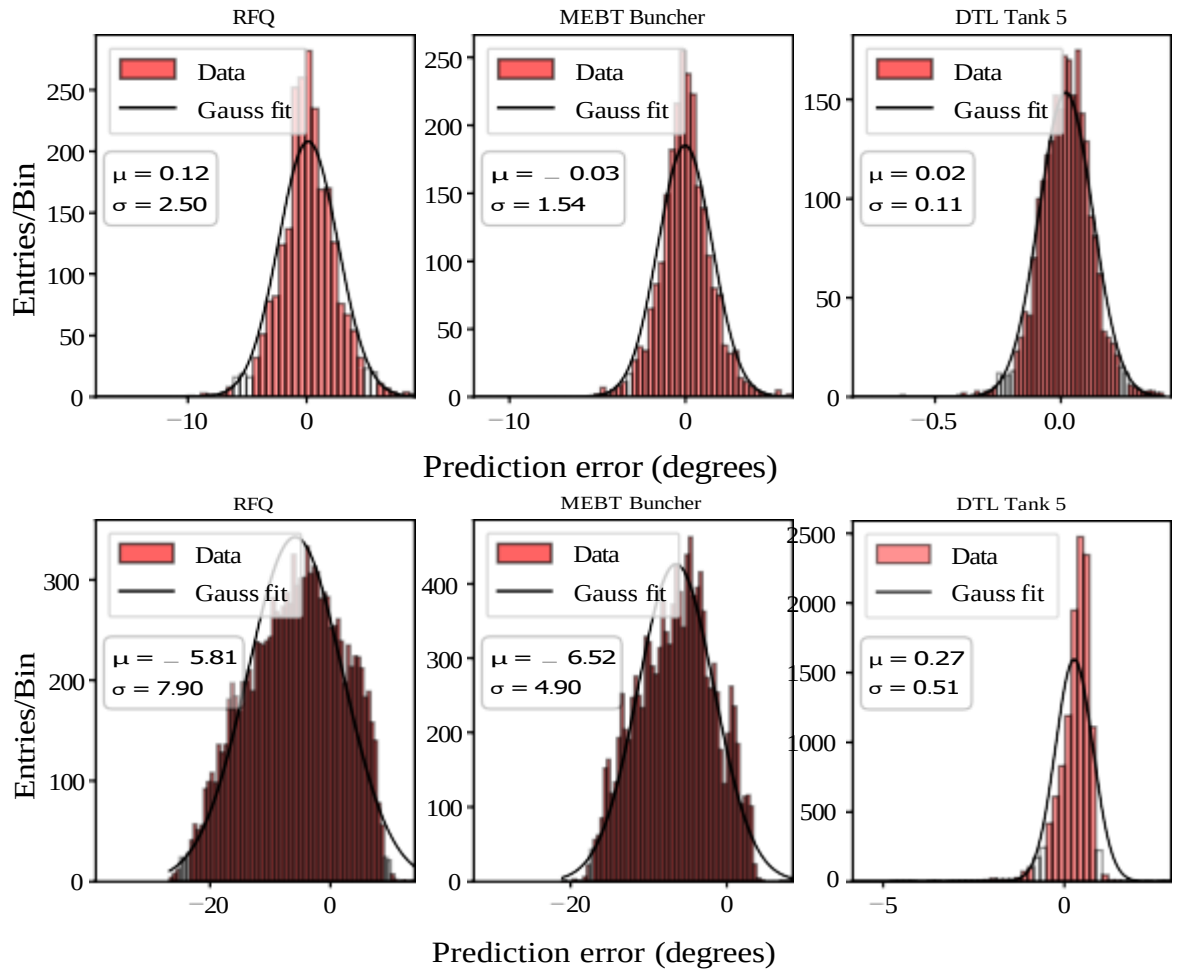

Figure 2: Network absolute prediction error. Gaussian mean and sigma inline. Top: data from same time period as training data, May 2022. Bottom: data from 3 weeks later.

Our next step is to pursue simulation to be able to generate a training dataset for our ML applications. By varying the initial energy and phase at the beginning of the Linac, as well as the RF phase set points and gradients for each of the 17 RF cavities, we will generate simulated data that is robust to changes both in the source and in the Linac and thus time-independent.

## LINAC INSTRUMENTATION

In parallel with our ML optimization efforts, we have been working on ensuring that diagnostics are robust and sufficient to be able to fully optimize the RF paremeters of the whole Linac with the goal of achieving constant energy and control longitudinal emittance. In addition to examining Linac BLM and BPM data as discussed above, we are working on two more diagnostic studies: bunch length measurements in the transition section and Linac output energy measurements.

### *Bunch Length Measurements*

To be able to better measure effects of upstream RF tuning on the matching of beam between DTL and SCL we recommissioned in 2021 a Bunch Length Detector (BLD) in the transition section. The BLD [4] measures the average bunch length of a train of bunches using secondary electrons generated as beam passes through a wire at high voltage. These electrons are synchronized them to the RF reference by a RF deflector. Adjusting the phase of the deflection allows to measure different longitudinal slices of the beam. Figure 4 shows two BLD measurements at normal operational conditions.

By fitting a Gaussian to the data we estimate the mean $\mu$ and the standard deviation $\sigma$. Figure 5 shows the mean and $\sigma$ as a function of DTL Tank 5 RF phase set point. Each point in these two plots is an average of two measurements at the same setting. The difference between the two measurements is taken as systematic error and is of the order of 0.2°@201 MHz. As expected, changing the phase shifts the signal mean, and the relationship appears to be linear. Bunch length appears flat within the uncertainty, suggesting the beam energy is largely unchanged. This result shows that BLD data is highly sensitive to upstream RF changes and can be used to improve beam matching at the transition section.

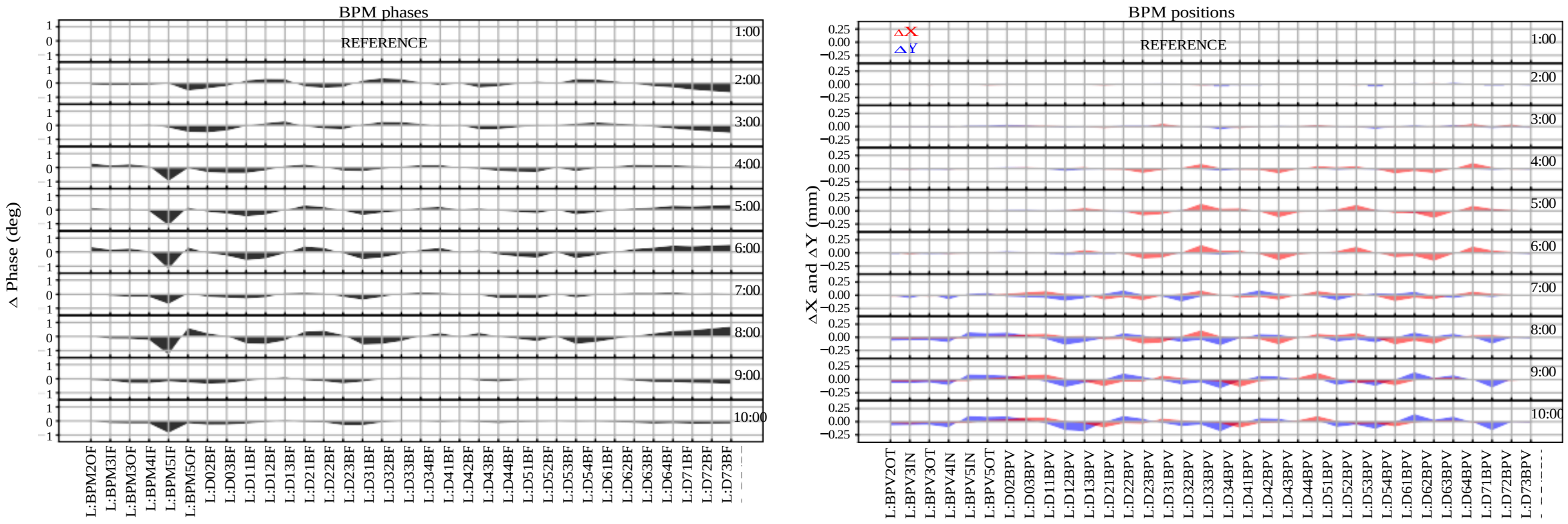

Figure 3: Evolution over 10 hours of BPM phases (left) and BPM horizontal and vertical positions (right).

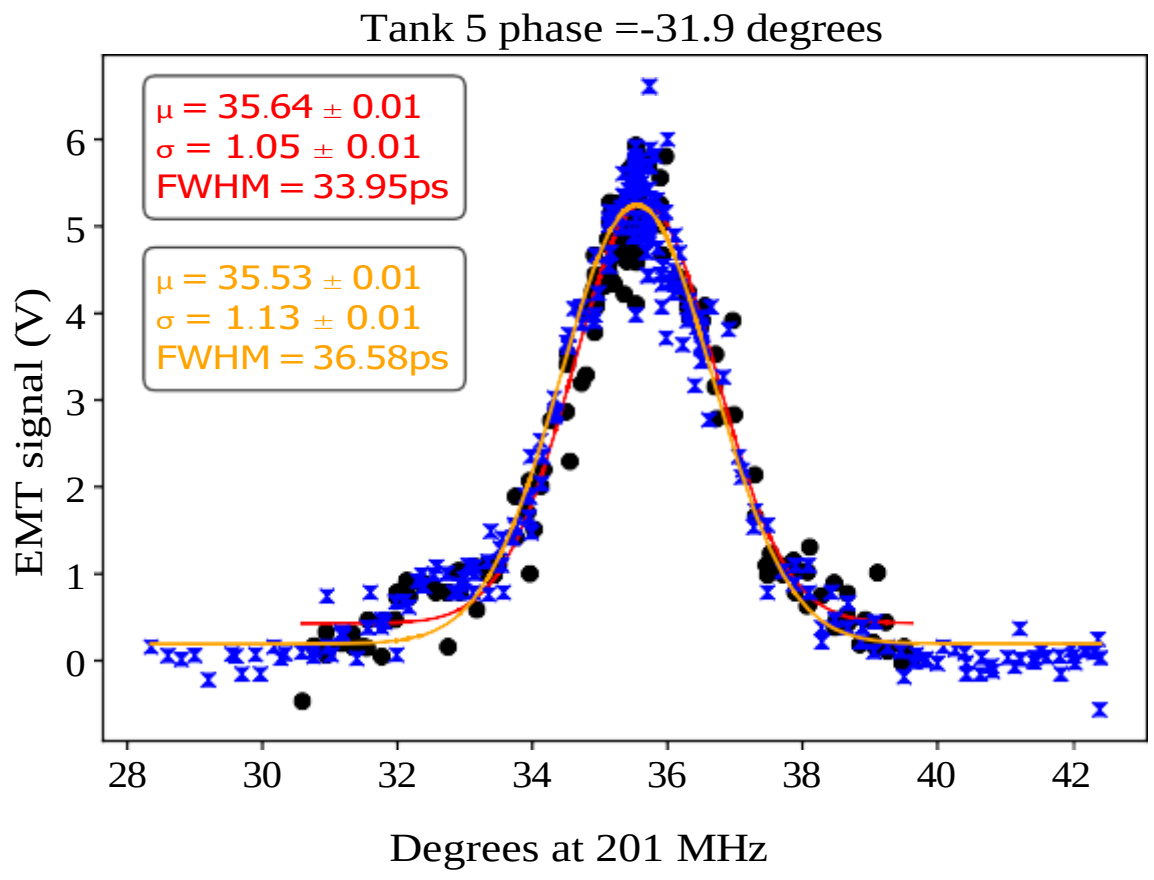

Figure 4: Typical longitudinal distributions from BLD.

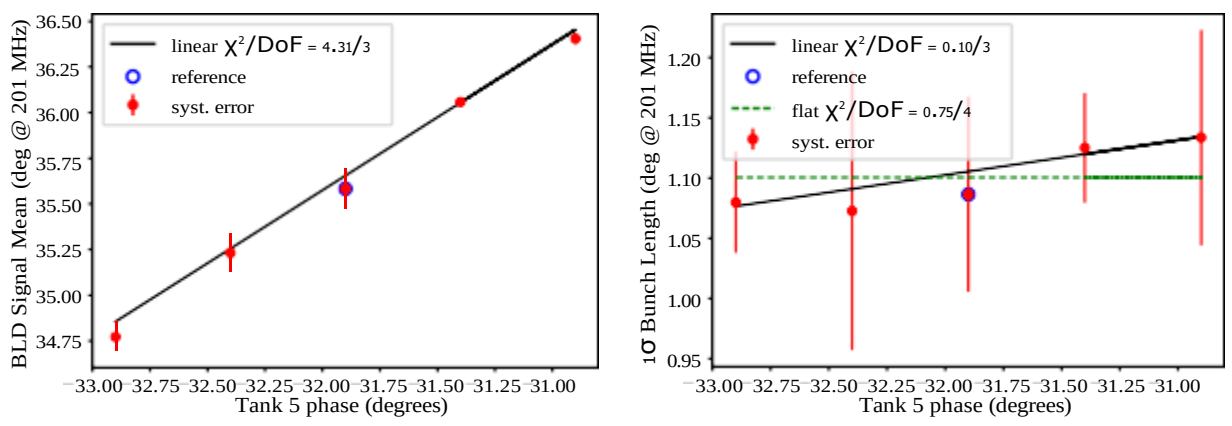

Figure 5: Mean and $\sigma$ of BLD distribution vs. Tank 5 phase.

### *Linac Energy Measurement*

Having constant energy throughout the Linac pulse is important for downstream machines. To ensure energy is stable we are developing three methods for measurement: Time of Flight (ToF) measurements with BPMs and longitudinal pickup monitors, and a method using dispersion. The pickups are ≈ 50 m apart, downstrea of the Linac. We recorded beam traces with a high-resolution (80 ps) oscilloscope and calculated the time offset between every two corresponding peaks (bunches) as shown in Fig. 6. The pulse length is 30$\mu$s, corresponding to ≈ 6000 bunches at the beam frequency of 201.25 MHz. The time offset was constant over the whole trace indicating that Linac pulse energy is constant within the scope resolution corresponding to ≈ 1 MeV.

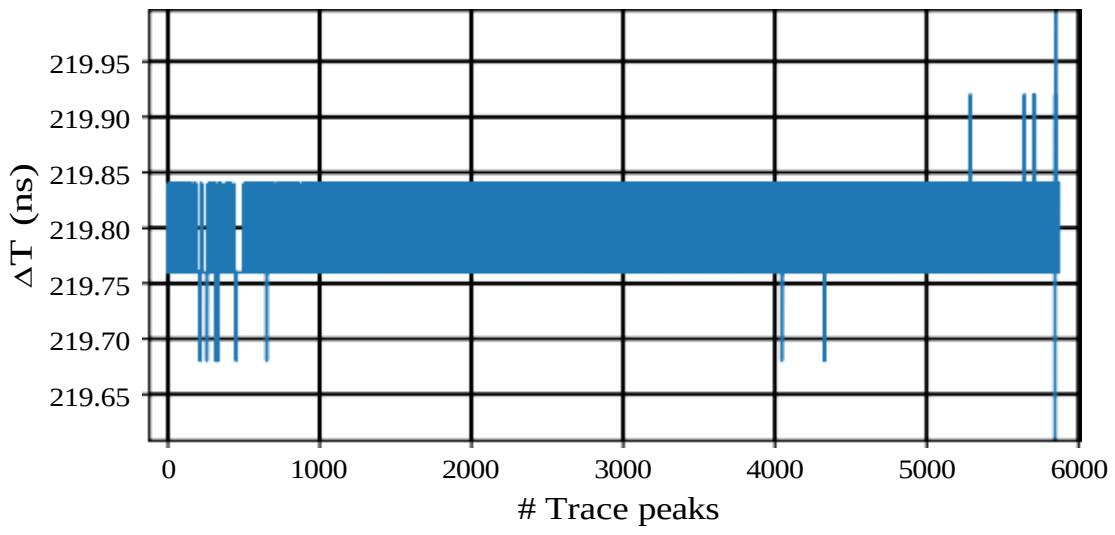

Figure 6: Time delay between corresponding peaks(bunches) in two beam pickups ≈ 50m apart downstream of Linac.

## CONCLUSION

We are exploring ML applications for predicting Linac RF parameters from diagnostics data to be used in a control scheme for automated optimization. Preliminary model is very promising, however account for ion source condition changes due to a lack of instrumentation. We plan to generate simulated data for NN training. Alongside ML efforts we are revisiting Linac diagnostics and developing new procedures to utilize data.

## ACKNOWLEDGEMENTS

This material is based upon work supported by the U.S. Department of Energy, Office of Science, Office of High Energy Physics, under contract number DE-AC02-07CH11359.